\documentclass[12pt]{article}
\usepackage{amsmath,amssymb,times,graphicx,mathdots}
\begin{document}
\pagestyle{plain}
\hsize = 6. in 				
\vsize = 8.5 in		
\hoffset = -0.3 in
\voffset = -0.5 in
\baselineskip = 0.26 in	

\def\vF{{\bf F}}
\def\vJ{{\bf J}}
\def\vX{{\bf X}}
\def\tf{{\widetilde{f}}}
\def\vv{{\bf{v}}}
\def\tv{{\widetilde{v}}}
\def\tu{{\widetilde{u}}}
\def\tp{{\widetilde{\rho}}}
\def\tR{{\widetilde{R}}}
\def\vx{\mbox{\boldmath$x$}}
\def\mQ{\mathbb{Q}}
\def\mM{\mathbb{M}}
\def\mR{\mathbb{R}}
\def\mP{\mathbb{P}}
\def\mcA{\mathcal{A}}
\def\wtQ{\widetilde{\mathcal{Q}}}

\title{A Hamiltonian-Entropy Production Connection
in the Skew-symmetric Part of a Stochastic Dynamics}

\author{Hong Qian\\[10pt]
Department of Applied Mathematics\\
University of Washington, Seattle\\
WA 98195-2420, U.S.A.}

\maketitle

\begin{abstract}
The infinitesimal transition probability operator
for a continuous-time discrete-state Markov process,
$\mathcal{Q}$, can be decomposed 
into a symmetric and a skew-symmetric
parts.  As recently shown for the case of diffusion processes, 
while the symmetric part corresponding to a gradient system
stands for a reversible Markov process, the skew-symmetric 
part, $\frac{d}{dt}u(t)=\mcA u$, is mathematically equivalent 
to a linear Hamiltonian dynamics with Hamiltonian $H=\frac{1}{2}u^T\big(\mcA^T\mcA)^{\frac{1}{2}}u$. It 
can also be transformed into a Schr\"{o}dinger-like equation
$\frac{d}{dt}u=i\mathcal{H}u$ where the ``Hamiltonian''
operator $\mathcal{H}=-i\mcA$ is Hermitian.  In fact, these 
two representations of a skew-symmetric dynamics emerge 
natually through singular-value and eigen-value decompositions,
respectively. The stationary probability of the 
Markov process can be expressed as $\|u^s_i\|^2$.  The 
motion can be viewed as ``harmonic'' since 
$\frac{d}{dt}\|u(t)-\vec{c}\|^2=0$ where 
$\vec{c}=(c,c,\cdots,c)$ with $c$ being a constant.  
More interestingly, we discover that 
\[
         \textrm{Tr}(\mcA^T\mcA)=\sum_{j,\ell=1}^n
      \frac{(q_{j\ell}\pi_\ell-q_{\ell j}\pi_j)^2}{\pi_j\pi_{\ell}},
\]
whose right-hand-side is intimately related to the entropy 
production rate of the Markov process in a nonequilibrium 
steady state with stationary distribution $\{\pi_j\}$.  
The physical implication of this intriguing connection 
between conservative Hamiltonian dynamics and dissipative
entropy production remains to be further explored.
\end{abstract}

\section{Introduction}

Linear operator theory and functional analysis became the 
centerpiece of quantum mechanics in the work of Dirac and 
von Neumann \cite{dirac,vonneumann,jordan}.  In 1930s,
Koopman, Birkhoff, von Neumann, and others have also 
developed a classical dynamical systems theory, including
several ergodic theorems, based on linear transformations in 
Hilbert space \cite{koopman,koopman_vonneumann,
birkhoff_koopman,khinchin}.  One can find this
approach to nonlinear dynamical systems in several excellent
treatises \cite{mackey_book,gaspard_book,mackey,mezic}.  
In a function space, a Koopman operator maps 
a function $\phi(x)$ to $U_t[\phi]=\phi(S_t(x))$ where 
$S_t(x)$ is the trajectory of an underlying dynamical system.  
Thus, it represents the dynamics in terms of a collection 
of arbitrary ``test functions'' defined on a moving 
coordinate system which follows a set of differential 
equations: ``$U_t[\phi]$ has at $x$ the value which 
$\phi$ has at the point $S_t(x)$ into which $x$ flows 
after the lapse of the time $t$'' \cite{koopman}.
This is the deterministic counterpart of Kolmogorov's 
backward equation; while the Perron-Frobenius operator 
corresponds to Kolmogorov's forward and Liouville 
equations \cite{mackey_book,gaspard_book}.

Koopman \cite{koopman} showed that
a Hamiltonian dynamics in a certain region of 
$\mathbb{R}^{2n}$ on a variety $H(q,p)=C$ of
points can be represented in terms
of a unitary transformation $U_t$ in an 
appropriate Hilbert space: 
$\big(U_t[\phi],U_t[\psi]\big)=(\phi,\psi)$.
Since $U_t$ is a family of one-parameter group,
it has an infinitesimal generator $\mathcal{G}$,
\begin{equation} 
   \left[\frac{\partial}{\partial t} U_t[\phi(x)]\right]_{t=0}
            = i\mathcal{G}_{\phi}(x).
\end{equation}
$\mathcal{G}$ is self-adjoint, or Hermitian:
$\big(\mathcal{G}_{\phi},\psi\big) =\big(\phi,\mathcal{G}_{\psi} \big)$.  

This paper studies the skew-symmetric linear operator derived 
from decomposition of continuous-time Markov processes \cite{qian_decomp}. It has been shown that for a stochastic
diffusion process \cite{qian_decomp}, the anti-symmetric part
corresponds to a hyperbolic system whose characteristic
lines follow a differential equation $\dot{x}=j(x)$ with
$\nabla\cdot\big(\rho(x)j(x)\big)=0$, where $\rho(x)$
is the stationary density of the diffusion process.    
Here we show that, as a finite-dimensional analogue of the
$\mathcal{G}$, the skew-symmetric part of a continuous-time, discrete-state Markov process in fact can
be further mathematically transformed into a Hamiltonian 
system with a symplectic strucuture.   This is the 
consequence of a skew-symmetric real operator 
$\mcA$ whose eigenvalues are pairs of imaginary numbers.

Based on the present result for systems with 
finite dimension, we suspect that an anti-symmetric 
operator $\mathcal{A}$ in an appropriate Hilbert space, 
derived from diffusion process decomposition, has a 
linear, Hamiltonian structure as well, with its Hamiltonian being
$\frac{1}{2}\big(\phi,\big(\mcA^T\mcA\big)^{\frac{1}{2}}\phi\big)$. Note that $(\phi,\mcA\phi)=0$; and in 
the dynamics defined by the anti-symmetric operator
$\frac{d}{dt}\phi(t) = \mcA\phi$, 
$\|\phi(t)\|^2=(\phi,\phi)$ is a 
constant of motion.  In fact, any operator $\mathcal{P}$ 
that commutes with $\mcA$, $\mcA\mathcal{P}=\mathcal{P}\mathcal{A}$, will have
\begin{equation}
  \frac{d}{dt}\big(\phi,\mathcal{P}\phi\big) = 
              \big(\mcA\phi,\mathcal{P}\phi\big)
                  +\big(\phi,\mathcal{P}\mcA\phi\big)
          = \big(\phi, [-\mcA\mathcal{P}
                         +\mathcal{P}\mcA]\phi\big) = 0.
\end{equation}

Mathematically, even for finite dimensional systems, the
present analysis is far from rigorous or complete; a full 
treatment remains to be developed.

\section{Decomposition and the skew-symmetric part}

\subsection{Dynamics and its different mathematical 
representations}

When a set of variables changing with time, we say 
there is a ``dynamics''.  Classical dynamics 
is customly represented by the time change of the
variables themselves, $x(t)$, in terms of a system of 
differential equations $\frac{d}{dt}x(t) = b(x)$.  In the
same vein, stochastic, Markov dynamics is represented
by $dx(t)=b(x)dt+\sigma dW(t)$, first appeared in the work
of Langevin, now widely known as a stochastic differential 
equation.  

The work in the 1930s by von Neumann, Koopman, and Birkhoff in 
USA \cite{koopman,koopman_vonneumann,birkhoff_koopman}, and 
Kolmogorov, Khinchin, and others in 
USSR \cite{kolmogorov,khinchin}, however, represents
a dynamics by a one-parameter family of linear operators 
in a function space.  In the case of Perron-Frobenius operator
\cite{mackey_book}, the corresponding Liouville equation and 
Kolmogorov forward equation are interpreted as the motion
of a {\em density function} for a collection of particles
following the classical dynamics.  The interpretation of the 
backward equation, or Koopman operator, on the other hand,
is an artitary ``test'' function in a moving coordinate system 
that follows the classical differential equation.   
These ``modern'' mathematical representations of dynamics
ultimately became the foundation of quantum mechanics
and stochastic processes.  Historically, it is worth 
pointing out that there was a ``direct personal 
correspondence between Schr\"{o}dinger and Kolmogorov at 
the time'' \cite{nagasawa}. 

	One of the insights from these earlier work is that 
abstract representations of dynamics, while might not
have simple or intuitive interpretations, can be powerful.
In fact, trajectory, forward, and backward are three 
different representations of a classical dynamics, 
deterministic or stochastic. While the relation
among an ordinary differential equation, its Liouville
equation and Koopman operator are unambiguously defined, the
relation between a stochastic differential equation
and its forward and backward equations involves It\={o},
Stratonovich, divergence-form, or other interpretations.  This
has been an important issue in the recent work of P.
Ao and his coworkers \cite{shi_ao,yuan_ao}.  It is also
noted that many studies on entropy productions of 
diffusion processes had also employed the divergence-form
elliptic operator \cite{esposito,maes,jqq}.

\subsection{Decomposition of a continuous-time Markov process}

Traditionally, a continuous-time, discrete-state Markov 
process (CTDS-MP) is characterized by its infinitesimal 
transition probability rate matrix, called Q-matrix, in 
terms of the master equation
\begin{equation}
       \frac{d}{dt}p(t) = \mathcal{Q} p(t),
\label{meq}
\end{equation}
in which column vector $p(t)=\{p_i(t)|i=1,2,\cdots,n\}$
represents the probability of a stochastic system in 
state $i$ at time $t$.  If the $p(t)$ is a matrix, 
Eq. \ref{meq} is also widely known as the Kolmogorov 
forward equation for the Markov process. It can also 
be understood in terms of a {\em Markov density matrix} 
representation.  See Appendix A.

The stochastic trajectory of a CTDS-MP
can be defined through a random time-changed Poisson
process in terms of multi-variate independent Poisson 
processes with unit rate \cite{kurtz}.  

Assuming the Markov process is irreducible and recurrent, 
let $\pi$ be its unique stationary distribution:
$\mathcal{Q}\pi = 0$.  We shall denote
diagonal matrix $\Pi = $ diag$(\pi_1,\pi_2,\cdots,\pi_n)$.
Then $\mathcal{Q}$ can be decomposed as a forward operator: 
$\mathcal{Q} = \mathcal{Q}_S+\mathcal{Q}_A$
with
\begin{equation}
   \mathcal{Q}_S = \frac{1}{2}\left(\mathcal{Q}
          +\Pi\mathcal{Q}^T\Pi^{-1}\right), \ \ 
    \mathcal{Q}_A = \frac{1}{2}\left(\mathcal{Q}
          -\Pi\mathcal{Q}^T\Pi^{-1}\right).
\end{equation} 
$\mathcal{Q}^T = \mathcal{Q}^T_S+\mathcal{Q}^T_A$
will be the corresponding decomposition for the backward
operator. (See Appendix B for a discussion on the diffusion
process.)

	Note a very important difference between this
decomposition for CTDS-MPs and for the decomposition for
diffusion processes \cite{qian_decomp}: The $\mathcal{Q}_A$ 
is no longer a proper Q-matrix; it has negative off-diagonal
elements.
 
	The dynamic equation (\ref{meq}) can be decomposed
accordingly:
\begin{equation}
      \frac{d}{dt} u(t)
         = \left(\Pi^{-\frac{1}{2}}\mathcal{Q}
           \Pi^{\frac{1}{2}}\right)
           \left(\Pi^{-\frac{1}{2}}p(t)\right)
         = \left(\mathcal{S}+\mathcal{A}\right) u(t)
\end{equation}
in which $u(t)=\Pi^{-\frac{1}{2}}p(t)$.  It has the
dimension of the square root of probability.

The symmetric part is well understood.  It corresponds
to a reversible Markov process \cite{jqq} with stationary
solution $u^s_i = \sqrt{\pi_i}$.  Thus, $|u^s_i|^2=\pi_i$
is the stationary probability distribution of the 
Markov process.  Extensive studies on the non-symmetric
part in term of a {\em circulation decomposition} theorem,
which characterizes along the path of a process in terms 
of reversible and rotational motions, can be found in \cite{jqq,qq82,qqg91,kalpazidou}, with applications to 
stationary flux analysis in physics and 
chemistry \cite{hill,wang1,wang2}. 

	The symmetric part as a $n$-dimensional linear system
\begin{equation}
        \frac{d}{dt} u(t) = \mathcal{S}u = -\nabla \Phi(u),
\end{equation}
has a gradient with potential 
$\Phi(u) = -\frac{1}{2}u^T\mathcal{S}u$.  The symmetric
matrix $(-\mathcal{S})$ is semi-positive. More
recently, the dynamics of the symmetric part has also
been shown as a gradient flow in an appropriate Riemannian 
manifold of probability distributions under a Wasserstein
metric \cite{chow,otto}.

\subsection{A linear Hamiltonian system}

	we now consider the skew-symmetric part as a 
$n$-dimensional linear system
\begin{equation}
        \frac{d}{dt} u(t) = \mcA
                    u,
\label{the_eq}
\end{equation}
in which real matrix $\mcA$ is skew-symmetric
$\mcA^T = -\mcA$.  Eigenvalues of $\mcA$ are 
pairs of imaginary conjugate numbers or zeros, say
$\pm i\lambda_1,\pm i\lambda_2,\cdots,\pm i\lambda_k$, 
$0,\cdots, 0$.  Therefore,
by an orthogonal matrix $B$, $B^{-1}=B^T$, a similarity
transformation relates $\mcA$ to\footnote{The orthogonal
matrix $B$ is real.  This will be shown in the next 
section.  In fact, $B=V^*$ in Sec. \ref{sec4.3}.} 
\begin{equation}
   \mathcal{H}_1= B\mathcal{A}B^T  
              = \left(\begin{array}{ccccccccc}
           0 & \lambda_1 & 0 & 0 & 0 & 0 & 0 & 0 & 0 \\
           -\lambda_1 & 0 & 0 & 0 & 0 & 0 & 0 & 0 & 0 \\
           0 & 0 & 0 & \lambda_2 & 0 & 0 & 0 & 0 & 0 \\
           0 & 0 & -\lambda_2 & 0 & 0 & 0 & 0 & 0 & 0 \\
           \vdots & \vdots & \vdots & \ddots & \ddots & \ddots & \vdots & \vdots & \vdots \\
           0 & 0 & 0 & 0 & 0 & 0 & \lambda_k & 0 & 0 \\
           0 & 0 & 0 & 0 & 0 & -\lambda_k & 0 & 0 & 0 \\
           0 & 0 & 0 & 0 & 0 & 0 & 0 & 0 & 0  \\
           0 & 0 & 0 & 0 & 0 & 0 & 0 & \ddots & 0  \\
           0 & 0 & 0 & 0 & 0 & 0 & 0 & 0 & 0
               \end{array}\right),
\label{large_ma}
\end{equation}
together with
\begin{equation}
    \big(u_1,u_2,\cdots,u_n\big)B^T = 
    \big(\underbrace{x_1,y_1,x_2,y_2,\cdots,x_k,y_k,\cdots}_n\big).
\label{u2xy}
\end{equation}
Note that the matrix transformation reveals canonical
pairs of variables $\big(x_i,y_i\big)$ which are hidden
under the original representation with $u$'s. 
The matrix $\mathcal{H}_1$ defines a linear Hamiltonian 
dynamical system (harmonic oscillator!) 
\begin{equation}
    \frac{dx_j}{dt} = \frac{\partial H}{\partial y_j}, \ \
    \frac{dy_j}{dt} = -\frac{\partial H}{\partial x_j},
\end{equation}
with Hamiltonian
\begin{equation}
    H(x_1,y_2,x_2,y_2,\cdots,x_k,y_k,\cdots)
      = \frac{1}{2}\sum_{j=1}^k\lambda_j
                              \left(x_j^2+y_j^2\right).
\label{hf}
\end{equation}

For the dynamics in (\ref{the_eq}), the Hamiltonian 
in (\ref{hf}) is $\frac{1}{2}u^T(\mcA^T\mcA)^{\frac{1}{2}}u$,  
where we introduce the notion,
\begin{eqnarray}
      (\mcA^T\mcA)^{\frac{1}{2}} &=&
    B^{-1}(B\mcA^TB^{-1}B\mcA B^{-1})^{\frac{1}{2}}B
\nonumber\\ 
          &=&  B^{-1}(\mathcal{H}_1^T
       \mathcal{H}_1)^{\frac{1}{2}}B.
\label{equation_12}
\end{eqnarray}
In Eq. \ref{equation_12}, $\big(\mathcal{H}_1^T\mathcal{H}_1\big)^{\frac{1}{2}}
= \Sigma =$ diag$(\lambda_1,\lambda_1,\lambda_2,\lambda_2,\cdots,\lambda_k,\lambda_k,0,\cdots,0)$ is the singular value matrix of $\mcA$ (see Sec. \ref{sec4.3}).  
Then following Eq. \ref{u2xy} we have,
\begin{equation}
     \frac{1}{2}u^T(\mcA^T\mcA)^{\frac{1}{2}}u 
         = H\big(u\big).
\label{H_u}
\end{equation}
Furthermore, one indeed has the conservation
\begin{eqnarray}
   \frac{d}{dt}u^T(\mcA^T\mcA)^{\frac{1}{2}}u
                &=&  u^T\left(\mcA^T(\mcA^T\mcA)^{\frac{1}{2}}
               +(\mcA^T\mcA)^{\frac{1}{2}}\mcA\right)u
\nonumber\\
         &=&  (Bu)^{T} \left( \mathcal{H}_1^T(\mathcal{H}_1^T\mathcal{H}_1)^{\frac{1}{2}}
      +(\mathcal{H}_1^T\mathcal{H}_1)^{\frac{1}{2}}\mathcal{H}_1 \right) (Bu)
\nonumber\\
     &=&  (Bu)^{T} \left( (\mathcal{H}_1^T
         \mathcal{H}_1)^{\frac{1}{2}}\left(\mathcal{H}_1^T
           +\mathcal{H}_1\right) \right) (Bu) \ = \ 0.
\end{eqnarray}

\subsection{A Schr\"{o}dinger-like equation}

Diagonalization of $\mathcal{H}_1$ requires
working with complex eigenvalues and eigenvectors.
Each $2\times 2$ block in (\ref{large_ma}) can be 
transformed
\begin{equation}
   \frac{1}{2}\left(\begin{array}{cc}
       -i & -1  \\ 1 & i \end{array}\right)
    \left(\begin{array}{cc}
       0 & \lambda  \\ -\lambda & 0 \end{array}\right)
      \left(\begin{array}{cc}
       i & 1 \\ -1 & -i \end{array}\right)
            =  i\left(\begin{array}{cc}
       \lambda & 0  \\ 0 & -\lambda \end{array}\right).
\label{EVD}
\end{equation}
Therefor one can denote $\mathcal{H}_1=i\mathcal{H}_2$ where
$\mathcal{H}_2$ is Hermitian.  Then $\mcA$ can be
written as $\mcA=B^T\mathcal{H}_1B
=iB^T\mathcal{H}_2B=i\mathcal{H}$, where $\mathcal{H}
=B^T\mathcal{H}_2B$ is Hermitian since $B$ is orthogonal.
The dynamics in (\ref{the_eq}) then has another, 
Schr\"{o}dinger-like, representation
\begin{equation}
        \frac{d}{dt} u(t) = i\mathcal{H}u.
\label{scheq}
\end{equation}
The Hamiltonian for (\ref{scheq}) is 
$\frac{1}{2}u^T\left(\mathcal{H}^2\right)^{\frac{1}{2}}u$.
Therefore, $\mathcal{H}$ can be legitimately called
a Hamiltonian operator. 

	From Eq. \ref{H_u}, it is also interesting to note that
\begin{equation}
      H(u) = \frac{1}{2} 
              \Big(u^T(\mcA^T\mcA)^{\frac{1}{2}}u\Big) =
            \textrm{Tr}\Big[\Sigma\big(B\hat{\rho}B^T\big)\Big],
\end{equation}
in which matrix $\hat{\rho}=uu^T$.  This representation 
is analogous to that of Heisenberg's in matrix mechanics
\cite{louisell}.

\section{Representations via decompositions of skew-symmetric matrix}

We now apply two widely used matrix analysis methods, 
eigenvalue decomposition (EVD) and singular-value 
decomposition (SVD), to a skew-symmetric
matrix \cite{xu}.  We shall show Schr\"{o}dinger-equation
like and Hamiltonian dynamics natually emerge in
these two representations, respectively.

\subsection{\boldmath{$2\times 2$} matrix}

First, let us consider skew-symmetric $2\times 2$ 
matrices in the general form
\[
         \lambda\left(\begin{array}{cc}
       0 & 1  \\ -1 & 0 \end{array}\right).
\]
It has an eigenvalue decomposition (EVD) (Eq. \ref{EVD}):        
\begin{equation} 
    \left(\begin{array}{cc}
       0 & 1  \\ -1 & 0 \end{array}\right)
            = \frac{1}{2} \left(\begin{array}{cc}
       1 & -i \\ i & -1 \end{array}\right)
            \left(\begin{array}{cc}
       i & 0  \\ 0 & -i\ \end{array}\right) 
        \left(\begin{array}{cc}
       1 & -i  \\ i & -1 \end{array}\right)
       = iB\Lambda B^*,
\label{eq28}
\end{equation}
and a singular-value decomposition (SVD):
\begin{equation}
       \left(\begin{array}{cc}
       0 & 1  \\ -1 & 0 \end{array}\right)
    =  \left(\begin{array}{cc}
       0 & 1 \\ 1 & 0 \end{array}\right)
            \left(\begin{array}{cc}
        1 & 0  \\ 0 & 1 \end{array}\right) 
            \left(\begin{array}{cc}
       -1 & 0  \\ 0 & 1 \end{array}\right)
          =U\Sigma V^T.
\label{2dSVD}
\end{equation}
Note that $U$ and $V$ in an SVD are not unique in general.  
However, for the $2\times 2$ problem, the general forms 
for orthogonal $U$ and $V$ are
\begin{equation}
     U =  \left(\begin{array}{cc}
                \cos\theta & \sin\theta \\
                -\sin\theta & \cos\theta \end{array}\right), 
         \  \  \ 
     V^T =  \left(\begin{array}{cc}
                \cos\phi & \sin\phi \\
                -\sin\phi & \cos\phi \end{array}\right) 
\end{equation}
with $\theta+\phi=\frac{\pi}{2}$.  Hence, $UV^T$
always gives the left-hand-side of Eq. \ref{2dSVD}.
Furthermore, $V^TU=UV^T$ and $VU^T=U^TV=-UV^T$.

\subsection{Eigenvalue decomposition (EVD)}

Let skew-symmetric, real matrix $\mcA$ has eigenvalues
and corresponding eigenvectors
$\mcA \vec{x}_{\ell} = i\lambda\vec{x}_{\ell}$,
where $\lambda_{\ell}$ are real and $\ell=1,2,\cdots,n$.
There is at least one $\lambda=0$;
and for even $n$, there are at least two zero eigenvalues.
Furthermore, each $i\lambda$ has a conjugate $-i\lambda$.
As a convention, we shall denote $\lambda_{2k}=-\lambda_{2k-1}
\le 0$.  Then
\[
         \mcA\vec{x}_{2k-1} = i\lambda_{2k-1}\vec{x}_{2k-1}, \ \ \
         \mcA\overline{\vec{x}_{2k-1}} 
          =-i\lambda_{2k-1}\overline{\vec{x}_{2k-1}}
          =i\lambda_{2k}\overline{\vec{x}_{2k-1}}.
\]
Therefore, $\vec{x}_{2k}= c\ \overline{\vec{x}_{2k-1}}$
where $c$ is a complex multiplier.
Note that $\vec{x}_{2k-1}$ and $\vec{x}_{2k}$ are orthonormal:
$\vec{x}_{2k-1}\cdot\overline{\vec{x}_{2k}}=0$.  Hence, 
$\vec{x}_{2k}\cdot\vec{x}_{2k}=0$, so is $\vec{x}_{2k-1}\cdot\vec{x}_{2k-1}=0$.  For example, these
are indeed the case for the column vectors of $B$ and
row vectors of $B^*$ in Eq. \ref{eq28}.

The EVD of $\mcA$ can then be written as 
\[
     \mcA = B\big(i\mathcal{H}\big)B^* =
   \left(\begin{array}{ccccccc}
       \big| & \big| & \big| &  \cdots & \big|   \\
       \vec{x}_1 & \vec{x}_2 & \vec{x}_3 & \cdots & \vec{x}_n \\
        \big| & \big| & \big| &  \cdots & \big|  \\
             \end{array}\right) \big( i\mathcal{H} \big)
              \overline{\left(\begin{array}{ccc}
               \textrm{---} & \vec{x}_1 &  \textrm{---} \\
               \textrm{---} & \vec{x}_2 &  \textrm{---} \\
                    \vdots  &  \vdots  & \vdots  \\
               \textrm{---} & \vec{x}_n &  \textrm{---}
             \end{array}\right)},
\]
in which 
\begin{equation}  
       \mathcal{H} = \left(\begin{array}{cccccccccc}
            \lambda_1 & 0 & 0 & 0 & \cdots &  0 & 0 & 0 \\
                0 &   -\lambda_1 & 0 & 0 & \cdots & 0 & 0 & 0 \\
                0 & 0 & \lambda_3 & 0 & \cdots & 0 & 0 & 0 \\
                0 & 0 & 0 & -\lambda_3 & \cdots & 0 & 0 & 0\\
                \vdots & \vdots & \vdots & \ddots & \ddots & 
                                 \ddots & \vdots & \vdots \\
            0 & 0 & 0 & 0 & \cdots & 0 & 0 & 0     
             \end{array}\right),
\end{equation}
and $\vec{x}_k\cdot\vec{x}_{\ell} = \delta_{k\ell}$.

The dynamics (\ref{the_eq}) in this
representation becomes $\frac{d}{dt}\varphi(t) 
= i\mathcal{H}\varphi$ in which $\varphi(t)=B^*u(t)$.

\subsection{Singular-value decomposition (SVD)}
\label{sec4.3}

The SVD of $\mcA$ gives
\[
     \mcA =
   \left(\begin{array}{ccccccc}
                       \\
       \big| & \big| & \big| &  \cdots & \big|   \\
       \vec{u}_1 & \vec{u}_2 & \vec{u}_3 & \cdots & \vec{u}_n \\
        \big| & \big| & \big| &  \cdots & \big|  \\
                        \\ 
             \end{array}\right)\Sigma 
            \left(\begin{array}{ccc}
               \textrm{---} & \vec{v}_1 &  \textrm{---} \\
               \textrm{---} & \vec{v}_2 &  \textrm{---} \\
                    \vdots  &  \vdots  & \vdots  \\
               \textrm{---} & \vec{v}_n &  \textrm{---}
             \end{array}\right), 
\]
in which the square diagonal
\begin{equation}  
       \Sigma = \left(\begin{array}{cccccccccc}
            |\lambda_1| & 0 & 0 & 0 & \cdots &  0 & 0 & 0 \\
                0 &   |\lambda_2| & 0 & 0 & \cdots & 0 & 0 & 0 \\
                0 & 0 & |\lambda_3| & 0 & \cdots & 0 & 0 & 0 \\
                0 & 0 & 0 & |\lambda_4| & \cdots & 0 & 0 & 0\\
                \vdots & \vdots & \vdots & \ddots & \ddots & 
                                 \ddots & \vdots & \vdots \\
            0 & 0 & 0 & 0 & \cdots & 0 & 0 & 0     
             \end{array}\right),
\end{equation} 
and $\mcA^T\mcA\vec{u}_{\ell}=|\lambda_{\ell}|^2\vec{u}_{\ell}$
with $|\lambda_{2k-1}|=|\lambda_{2k}|$.  Similarly,
$\mcA^T\mcA\vec{v}_{\ell}=|\lambda_{\ell}|^2\vec{v}_{\ell}$.
Both $U$ and $V$ are themselves orthogonal matrices.
Furthermore, $\vec{v}_{2k-1}=\vec{u}_{2k}$ and
$\vec{v}_{2k}=-\vec{u}_{2k-1}$.  Therefore, we have
\[
         \left(\begin{array}{ccc}
               \textrm{---} & \vec{v}_1 &  \textrm{---} \\
               \textrm{---} & \vec{v}_2 &  \textrm{---} \\
                    \vdots  &  \vdots  & \vdots  \\
               \textrm{---} & \vec{v}_n &  \textrm{---}
             \end{array}\right)
    \left(\begin{array}{ccccccc}
                       \\
       \big| & \big| & \big| &  \cdots & \big|   \\
       \vec{u}_1 & \vec{u}_2 & \vec{u}_3 & \cdots & \vec{u}_n \\
        \big| & \big| & \big| &  \cdots & \big|  \\
                        \\ 
             \end{array}\right)
\]
\begin{equation}
=   \left(\begin{array}{cccccccccc}
           0 & 1 & 0 & 0 & \cdots & 0 & 0 & 0 & 0 & 0 \\
           -1 & 0 & 0 & 0 & \cdots & 0 & 0 & 0 & 0 & 0 \\
           0 & 0 & 0 & 1 & \cdots & 0 & 0 & 0 & 0 & 0 \\
           0 & 0 & -1 & 0 & \cdots & 0 & 0 & 0 & 0 & 0 \\
           \vdots & \ddots & \ddots & \ddots & \ddots & \ddots
                   & \vdots & \vdots & \vdots & \vdots \\
            0 & \cdots & 0 & 0 & 0 & 0 & 1 & 0 & 0 & 0  \\ 
            0 & \cdots & 0 & 0 & 0 & -1 & 0 & 0 & 0 & 0  \\
            0 & \cdots & 0 & 0 & 0 & 0 & 0 & 0 & 0 & 0 \\
            \vdots &  & \vdots & \vdots& \vdots& 
            \vdots & \vdots& \vdots & \ddots & \vdots \\
            0 & \cdots & 0 & 0 & 0 & 0 & 0 & 0 & 0 & 0 
             \end{array}\right) \triangleq
             \widetilde{\mathcal{H}}_1.
\end{equation}

The dynamics (\ref{the_eq}) in this
representation then becomes 
\begin{equation}
      \frac{d}{dt}\xi(t) = V^*U\Sigma\ \xi 
               = \widetilde{\mathcal{H}}_1\Sigma\ \xi,
\end{equation}
in which $\xi(t)=V^*u(t)$. $\widetilde{\mathcal{H}}_1$
reveals a symplectic structure of the dynamics.
$\widetilde{\mathcal{H}}_1\Sigma=\mathcal{H}_1$ in
Eq. \ref{large_ma}.

\subsection{Relationships between \boldmath$\vec{x}$ and $\vec{u}$}

While vector $\vec{x}_{\ell}$ are complex, vector
$\vec{u}_{\ell}$ are real.  We have
$\mcA\vec{x}_{\ell} = i\lambda_{\ell}\vec{x}_{\ell}$,
$\mcA^T\mcA\vec{x}_{\ell} = \lambda_{\ell}^2\vec{x}_{\ell}$
Noting $\lambda_{2k-1}^2=\lambda_{2k}^2$, we therefore have, 
\begin{subequations}
\begin{equation}
         \vec{u}_{2k-1} = \alpha_{11} \vec{x}_{2k-1}
                      +\alpha_{12}\vec{x}_{2k},
\end{equation}
\begin{equation}
         \vec{u}_{2k} = \alpha_{21} \vec{x}_{2k-1}
                          +\alpha_{22}\vec{x}_{2k}.
\end{equation}
\end{subequations}
Accoding to the example in (\ref{eq28}) and (\ref{2dSVD}),
\[
   \left(\begin{array}{cc}
                 \alpha_{11} & \alpha_{12} \\
                 \alpha_{21} & \alpha_{22}
               \end{array}\right) =
    \left(\begin{array}{cc}
                 -\frac{i}{2} & -\frac{1}{2} \\[6pt]
                 \frac{1}{2} & \frac{i}{2}
               \end{array}\right).
\]

\section{Trace, Hamiltonian, and entropy production}

	In terms of the original Markov process with
infinitesimal transition rate matrix $\mathcal{Q}$, 
the elements of matrix 
$\mathcal{A}=\Pi^{-\frac{1}{2}}\mathcal{Q}\Pi^{\frac{1}{2}}
-\Pi^{\frac{1}{2}}\mathcal{Q}^T\Pi^{-\frac{1}{2}}$ is:
\begin{equation}
       a_{ij} = \frac{q_{ij}\pi_j-q_{ji}\pi_i}
               {\sqrt{\pi_i\pi_j}}.
\end{equation}
It turns out that the trace of $\mathcal{A}^T\mathcal{A}$
is:
\begin{equation}
    \textrm{Tr}\big(\mathcal{A}^T\mathcal{A}\big)
       = \sum_{i,j=1}^n a_{ij}^2
       = \sum_{i,j=1}^n \frac{(q_{ij}\pi_j-q_{ji}\pi_i)^2}
               {\pi_i\pi_j}.
\label{eq22}
\end{equation}
Therefore, we have 
\begin{equation}
    \textrm{Tr}\big(\mathcal{A}^T\mathcal{A}\big)
            = \sum_{\ell=1}^n
                  \|\lambda_{\ell}\|^2 = \sum_{i,j=1}^n a^2_{ij}.  
\end{equation}

The implication of the mathematical equation in (\ref{eq22}) 
is very intriguing since according to the nonequilibrium 
steady state (NESS) theory of a Markov process, its entropy
production \cite{jqq,zqq,gqq} is 
\begin{equation}
   e_p = \sum_{i>j} \left(q_{ij}\pi_j-q_{ji}\pi_i\right)
         \ln\left(\frac{q_{ij}\pi_j}
                      {q_{ji}\pi_i}\right). 
\end{equation}
In particular, when a system is near an equilibrium, 
$q_{ij}\pi_j \simeq q_{ji}\pi_i$, then 
\begin{equation}
    e_p \simeq \frac{1}{2}\sum_{i,j=1}^n \frac{\big(q_{ij}\pi_j-
             q_{ji}\pi_i\big)^2}{q_{ji}\pi_i}.
\end{equation}

\section{Discussion}

Since 1930s, it has been known that three different 
types of dynamics, classical deterministic,
quantum, and stochastic dynamics, can all be 
represented in terms of linear operators in function
space.  While Schr\"{o}dinger, Dirac,
von Neumann's quantum mechanics, and Kolmogorov's 
forward and backward equations for stochastic processes 
are widely known, Koopman's powerful approach to 
classical dynamics has been mainly limited in 
mathematical literature.  While Newton and Hamiltonian's 
classical dynamics are conservative for mechanical
energy, Fourier's analytical theory of heat, together with 
the heat equation which turns out to the Kolmogorov 
equation for pure Brownian motion, has been a canonical 
example of dissipative systems.  
 
There is a growing interest in treating stochastic dynamics and 
statistical thermodynamics in a unified framework 
\cite{spohn,misra_prigogine,mackey,bedeaux,rubi,ao_05,
qian_jpc,ao_08,qian_jsp}. 
Recently, a decomposition of general stochastic diffusion 
dynamics in function space into symmetric and anti-symmetric
parts has shown that the former generalizes precisely Fourier's 
heat equation, while the latter generalizes Newtonian 
conservative dynamics \cite{qian_decomp}.  Furthermore, the
dynamics decomposition fits perfectly with a recently
discovered free energy balance equation: The symmetric part
has ``free energy decreasing $=$ entropy production'', as
was known to Helmholtz and Gibbs, and the anti-symmetric 
dynamics has {\em free} energy conservation.
 
	A mathematical investigation of 
anti-symmetric dynamics in a Hilbert space will be desirable.  
In the present work, we seek insights on the anti-symmetric, 
or skew-symmetric dynamics from finite dimensional systems.  
It is shown that both Hamiltonian representation and 
Schr\"{o}dinger-like representation natually emergy in
the singular-value decomposition and eigen-value decomposition
of an skew-symmetric matrix.  Finally, we discover an
intriguing connection between the Hamiltonian  for
the skew-symmetric dynamics and the entropy production rate
of the original irreversible Markov process. 
This observation calls for re-thinking of the nature of 
dissipation and time reversibility \cite{qian_decomp}.

\vskip 0.5cm\noindent

{\bf ACKNOWLEDGEMENTS.} I thank Ping Ao, Zhen-qing Chen,
Rafael De La Madrid, Hao Ge, Eli Shlizerman, Jin Wang,
and Wen-An Yong for helpful discussions.

\section{Appendices}

\subsection{Appendix A: Density matrix for a Markov process}

The solution to Eq. \ref{meq} can be formally written as
\begin{equation}
    p(t) = e^{\mathcal{Q}t}
             p(0).
\end{equation}
One can introduce a {\em Markov density matrix}:
\begin{equation}
   \rho_M(t) = e^{\mathcal{Q}t}
             \big|p(0) \rangle\langle 1\big|
\end{equation}
in which $|p(0)\rangle$ is an $n\times 1$ matrix,
i.e., a column vector, and $\langle 1|$ is 
a row vector consisting of 1s. Therefore, 
$\rho_M(t)$ is a matrix with Tr$(\rho_M)$ $=$
$\langle 1|p(t)\rangle=1$. Furthermore,
\begin{equation}
   \rho_M^2 = e^{\mathcal{Q}t}
             \big|p(0)\rangle\langle 1\big| = \rho_M.
\end{equation}
Then we have $\rho_M(t)$ satisfying the same dynamic 
equation as the Kolmogorov forward equation
\begin{equation}
  \frac{d}{dt}\rho(t) = \mathcal{Q}\rho,
\label{eq00096}
\end{equation}
with a difference in the initial data $\rho(0)$:
It gives a transition probability matrix if 
$\rho(0)=\mathbb{I}$; and it gives a density
matrix if $\rho_M(0)= \big|p(0) \rangle\langle 1\big|$.

	Similarly, a density matrix approach can be formulated
for Eq. \ref{scheq}.  It yields
\begin{equation}
   \frac{d}{dt}\rho(t) = i\big[\mathcal{H}\rho-
              \rho\mathcal{H}\big].
\end{equation}

\subsection{Appendix B}

The recently introduced ``canonical conservative 
dynamics'' \cite{qian_decomp} had been discussed 
in \cite{koopman}, in which the inner product in a
Hilbert space is defined with $\rho(x)$ as a weight 
\cite{jqq}; $\rho(x)$ being a positive, single-valued,
analytic function on $\mathbb{R}^n$. 
In the contrary, the weight used in
\cite{qian_decomp} is $\rho^{-1}(x)$.
This difference can be seen in the
matrix theory: Both $\Pi\mathcal{Q}$ and 
$\mathcal{Q}\Pi^{-1}$ are symmetric for reversible 
Markov process with generator $\mathcal{Q}$, but
only $\Pi^{-\frac{1}{2}}\mathcal{Q}\Pi^{\frac{1}{2}}$
tranforms the master equation into a gradient system.  
In fact, Koopman operator and Perron-Frobenius
operator belong to two different Hilbert spaces
with inner products
\begin{equation}
   \big(\phi,\psi\big)_{K} =
    \int_{\mathbb{R}^n}\rho(x)\phi(x)\psi(x)dx, 
\end{equation}
and
\begin{equation}
   \big(\phi,\psi\big)_{PF} =
         \int_{\mathbb{R}^n}\rho^{-1}(x)\phi(x)\psi(x)dx, 
\end{equation}
respectively.  Therefore, the symmetric operator
in the Koopman (backward) space is
\begin{equation}
     \mathcal{L}^*_S[u] = \frac{1}{2}\left(
     \mathcal{L}^*[u] + \rho^{-1}\mathcal{L}[\rho u] \right)
      = \nabla\cdot \big(A(x)\nabla u\big)
        +\big(\nabla\ln\rho\big)A(x)(\nabla u),
\end{equation}
while in the Perron-Frobenius (forward) space 
is \cite{qian_decomp}
\begin{equation}
    \mathcal{L}_S[u] = \frac{1}{2}\left(\mathcal{L}[u]
             +\rho\mathcal{L}^*\left[\rho^{-1}u\right]\right)
       = \nabla\cdot A(x)\big(\nabla u -\left(\nabla\ln\rho\right)u\big).
\end{equation}
Indeed, $\mathcal{L}_S$ and $\mathcal{L}_S^*$ correspond
to the Kolmogorov forward and backward equations of a
reversible diffusion. 

\subsection{Appendix C: Examples of skew-symmetric dynamics}

{\bf\em Simple continous skew-symmetric operator: Schr\"{o}dinger
 equation.}  Let us consider the Schr\"{o}dinger equation in free
space without potential:
\begin{equation}
   \frac{\partial\psi(x,t)}{\partial t} = 
           i\left(\frac{\hbar}{2m}\right)\frac{\partial^2}{\partial x^2}\psi(x,t),
\label{scheq}
\end{equation}
with initial value
\begin{equation}
        \psi(x,0) =  \frac{1}{\big(2\pi\sigma_0^2\big)^{\frac{1}{4}}}
                 e^{-x^2/(4\sigma_0^2)}.
\end{equation}
The solution is
\begin{equation}
    \psi(x,t) =  \big(8\pi\sigma_0^2\big)^{\frac{1}{4}} \sqrt{\frac{1}
       {4\pi\sigma_0^2+i2\pi\hbar t/m}}
         \exp\left[-\frac{x^2}{4\sigma_0^2+i2\hbar t/m}\right].
\end{equation}
Then,
\begin{equation}
    \|\psi(x,t)\|^2 = \sqrt{\frac{8\sigma_0^2/\pi}
        {\big(4\sigma_0^2\big)^2+\big(2\hbar t/m\big)^2}}
         \exp\left[-\left(\frac{8\sigma_0^2x^2}
           {\big(4\sigma_0^2\big)^2+\big(2\hbar t/m\big)^2}\right)
                \right].
\label{gaussiandistr}
\end{equation}
This is a normalized Gaussian distribution with variance
\begin{equation}
                  \sigma^2(t) = 
           \sigma_0^2+\left(\frac{\hbar t}{2m\sigma_0}
                                \right)^2.
\end{equation}
The variance is not growing with $t$, but $t^2$:
This is not diffusion.  The ``apparent velocity'' is 
$\pm\hbar/(2m\sigma_0)$.  The factor $\sigma_0$
means Heisenberg's uncertainty principle:  If the
initial data has an ``accuracy'' of $\sigma_0$, 
then the uncertainty in the velocity is $\hbar/\sigma_0$.

The solution to Eqn. (\ref{scheq})  has a 
Fouier representation in time:
\[
        \psi(x,t) = \int_0^{\infty} \Big\{
                  \left(\alpha_1e^{\sqrt{\frac{2m\omega}{\hbar}}x}
     +\beta_1 e^{-\sqrt{\frac{2m\omega}{\hbar}}x}\right)e^{i\omega t}
              \hspace{4in}
\]
\vskip -1cm
\begin{eqnarray}
       &+&  \left(\alpha_2\cos\sqrt{2m\omega/\hbar}\ x +
                \beta_2\sin\sqrt{2m\omega/\hbar}\ x\right)e^{-i\omega t}
              \Big\}\  d\omega
\nonumber\\[6pt]
      &=& \int_0^{\infty} \left\{
               \left[\alpha_1e^{\sqrt{\frac{2m\omega}{\hbar}}x}
     +\beta_1 e^{-\sqrt{\frac{2m\omega}{\hbar}}x}
      +\alpha_2\cos\sqrt{\frac{2m\omega}{\hbar}}\ x +
       \beta_2\sin\sqrt{\frac{2m\omega}{\hbar}}\ x \right]\cos\omega t
                     \right.
\nonumber\\
    &+&  \left. i\left[\alpha_1e^{\sqrt{\frac{2m\omega}{\hbar}}x}
     +\beta_1 e^{-\sqrt{\frac{2m\omega}{\hbar}}x}
      -\alpha_2\cos\sqrt{\frac{2m\omega}{\hbar}}\ x -
       \beta_2\sin\sqrt{\frac{2m\omega}{\hbar}}\ x \right]\sin\omega t
                  \right\} d\omega
\nonumber\\
\end{eqnarray}
Therefore, if $\psi(x,0)$ is an even, real-valued function of 
$x$,\footnote{More precisely, we only know the $\|\psi(x,0)\|^2$, but
this does not uniquely specify the $\psi(x,0)$.  Thererfore,
there is a uncertainty in the initial value for $\psi(x,t)$.   Different
choices here lead to different behaviour in the following
dynamics for $t>0$.
}
 we keep
only the $\alpha_2(\omega)$ term:
\begin{equation}
        \psi(x,t) = \Big(8\pi\sigma_0^2\Big)^{\frac{1}{4}}
                   \int_{-\infty}^{\infty}  e^{-\sigma_0^2\omega^2}
                       \cos\sqrt{\frac{2m\omega}{\hbar}} x\
                       \Big( \cos\omega t-i\sin\omega t\Big)\ d\omega.
\end{equation}
Then
\[
        \|\psi(x,t)\|^2 =   \Big(8\pi\sigma_0^2\Big)^{\frac{1}{2}}
               \int_0^{\infty}\int_0^{\infty} 
                    e^{-\sigma_0^2(\omega^2+\omega'^2)}
                       \cos\sqrt{\frac{2m\omega}{\hbar}}x\
                        \cos\sqrt{\frac{2m\omega'}{\hbar}} x
\]
\begin{equation}
              \Big\{ \cos\omega t \cos\omega' t   + 	
                        \sin\omega t \sin\omega' t 
               \Big\}\  d\omega\ d\omega'.
\label{eqno8}
\end{equation}

{\bf\em A discrete skew-symmetric dynamics.}  Consider
anti-symmetric dynamics
\begin{equation}
                    \frac{dx}{dt} = A x
\end{equation}
with
\begin{equation}
        A = \left(\begin{array}{cccc}   0  & -\lambda_1 & 0 & 0 \\ 
                                                   \lambda_1 & 0 & 0 & 0 \\
                                                   0 & 0 & 0 & -\lambda_2 \\
                                                   0 & 0 & \lambda_2 & 0 
                                                \end{array}\right).
\end{equation}
Then the solution to equation is
\begin{equation}
                  x(t) = e^{A t} x(0) \ \textrm{ in which } \
       e^{At} = \left(\begin{array}{cccc}   
               \cos \lambda_1 t  &  -\sin \lambda_1 t  & 0 & 0\\ 
               \sin \lambda_1 t   & \cos \lambda_1 t & 0 & 0 \\
               0 & 0 & \cos\lambda_2 t & -\sin\lambda_2 t \\
               0 & 0 & \sin\lambda_2 t & \cos\lambda_2 t   \end{array}\right).
\end{equation}
Therefore, if $x(0)=\left(\frac{1}{\sqrt{2}},0,\frac{1}{\sqrt{2}},0\right)^T$, then
\begin{equation}
                    \left(\begin{array}{c}
                            \|x_1(t)\|^2 \\[5pt]  \|x_2(t)\|^2 \\[5pt] 
                            \|x_3(t)\|^2 \\[5pt]  \|x_4(t)\|^2 \end{array}\right)
                      = \frac{1}{2}\left(\begin{array}{c}
                           \cos^2\lambda_1 t \\[5pt] \sin^2\lambda_1 t \\[5pt]
                           \cos^2\lambda_2 t \\[5pt] \sin^2\lambda_2 t 
                                        \end{array}\right). 
\end{equation}
However, if  $x(0)=\left(\frac{1}{2},\frac{1}{2},\frac{1}{\sqrt{2}},0\right)^T$, then
\begin{equation}
                    \left(\begin{array}{c}
                            \|x_1(t)\|^2 \\[5pt]  \|x_2(t)\|^2 \\[5pt] 
                            \|x_3(t)\|^2 \\[5pt]  \|x_4(t)\|^2 \end{array}\right)
                      = \left(\begin{array}{c}
                          \frac{1-\sin 2\lambda_1 t}{4} \\[5pt] 
                          \frac{1+\sin 2\lambda_1 t}{4} \\[6pt]
                           \frac{\cos^2\lambda_2 t}{2} \\[5pt] 
                           \frac{\sin^2\lambda_2 t}{2} 
                                        \end{array}\right). 
\label{eqno13}
\end{equation}
If $\lambda_1$ and $\lambda_2$ are non-commensurate, then
the $x(t)$ is not periodic.   If  dimension of $A$ is odd, then
it has a zero eigenvalue. 

	The finite dimensional Eqn. (\ref{eqno13}) should be compared
with the infinite dimensional Eqn. (\ref{eqno8}).  Heisenberg's uncertainty 
in the continuous dynamics is a consequence of infinite number of 
non-commensurate frequencies.

\end{document}